\documentclass[twoside,twocolumn,9pt]{article}
\usepackage{extsizes}
\usepackage{bm}
\usepackage[super,sort&compress,comma]{natbib} 
\usepackage[version=3]{mhchem}
\usepackage[left=1.5cm, right=1.5cm, top=1.785cm, bottom=2.0cm]{geometry}
\usepackage{balance}
\usepackage{mathptmx}
\usepackage{sectsty}
\usepackage{graphicx} 
\usepackage{lastpage}
\usepackage[format=plain,justification=justified,singlelinecheck=false,font={stretch=1.125,small,sf},labelfont=bf,labelsep=space]{caption}
\usepackage{accents}

\usepackage{float}
\usepackage{fancyhdr}
\usepackage{fnpos}
\usepackage[english]{babel}
\addto{\captionsenglish}{%
  
}
\usepackage{array}
\usepackage{droidsans}
\usepackage{charter}
\usepackage[T1]{fontenc}
\usepackage[usenames,dvipsnames]{xcolor}
\usepackage{setspace}
\usepackage[compact]{titlesec}
\usepackage{hyperref}
\usepackage{amsmath,amssymb}

\definecolor{cream}{RGB}{222,217,201}

\begin{document}

\pagestyle{fancy}
\thispagestyle{plain}
\fancypagestyle{plain}{
\renewcommand{\headrulewidth}{0pt}
}

\makeFNbottom
\makeatletter
\renewcommand\LARGE{\@setfontsize\LARGE{15pt}{17}}
\renewcommand\Large{\@setfontsize\Large{12pt}{14}}
\renewcommand\large{\@setfontsize\large{10pt}{12}}
\renewcommand\footnotesize{\@setfontsize\footnotesize{7pt}{10}}
\makeatother

\renewcommand{\thefootnote}{\fnsymbol{footnote}}
\renewcommand\footnoterule{\vspace*{1pt}%
\color{cream}\hrule width 3.5in height 0.4pt \color{black}\vspace*{5pt}} 
\setcounter{secnumdepth}{5}

\makeatletter 
\renewcommand\@biblabel[1]{#1}            
\renewcommand\@makefntext[1]%
{\noindent\makebox[0pt][r]{\@thefnmark\,}#1}
\makeatother 
\renewcommand{\figurename}{\small{Fig.}~}
\sectionfont{\sffamily\Large}
\subsectionfont{\normalsize}
\subsubsectionfont{\bf}
\setstretch{1.125} 
\setlength{\skip\footins}{0.8cm}
\setlength{\footnotesep}{0.25cm}
\setlength{\jot}{10pt}
\titlespacing*{\section}{0pt}{4pt}{4pt}
\titlespacing*{\subsection}{0pt}{15pt}{1pt}

\fancyfoot{}
\fancyfoot[LO,RE]{\vspace{-7.1pt}\includegraphics[height=9pt]{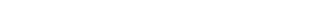}}
\fancyfoot[CO]{\vspace{-7.1pt}\hspace{13.2cm}\includegraphics{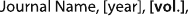}}
\fancyfoot[CE]{\vspace{-7.2pt}\hspace{-14.2cm}\includegraphics{head_foot/RF}}
\fancyfoot[RO]{\footnotesize{\sffamily{1--\pageref{LastPage} ~\textbar  \hspace{2pt}\thepage}}}
\fancyfoot[LE]{\footnotesize{\sffamily{\thepage~\textbar\hspace{3.45cm} 1--\pageref{LastPage}}}}
\fancyhead{}
\renewcommand{\headrulewidth}{0pt} 
\renewcommand{\footrulewidth}{0pt}
\setlength{\arrayrulewidth}{1pt}
\setlength{\columnsep}{6.5mm}
\setlength\bibsep{1pt}

\makeatletter 
\newlength{\figrulesep} 
\setlength{\figrulesep}{0.5\textfloatsep} 

\newcommand{\topfigrule}{\vspace*{-1pt}%
\noindent{\color{cream}\rule[-\figrulesep]{\columnwidth}{1.5pt}} }

\newcommand{\botfigrule}{\vspace*{-2pt}%
\noindent{\color{cream}\rule[\figrulesep]{\columnwidth}{1.5pt}} }

\newcommand{\dblfigrule}{\vspace*{-1pt}%
\noindent{\color{cream}\rule[-\figrulesep]{\textwidth}{1.5pt}} }

\makeatother

\twocolumn[
  \begin{@twocolumnfalse}
{\includegraphics[height=30pt]{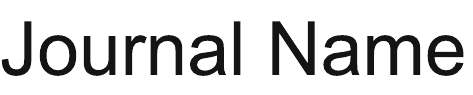}\hfill\raisebox{0pt}[0pt][0pt]{\includegraphics[height=55pt]{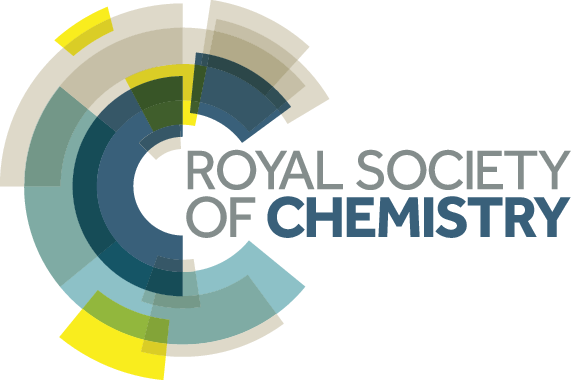}}\\[1ex]
\includegraphics[width=18.5cm]{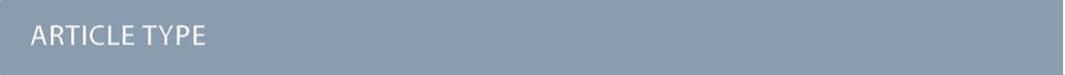}}\par
\vspace{1em}
\sffamily
\begin{tabular}{m{4.5cm} p{13.5cm} }

\includegraphics{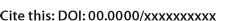} & \noindent\LARGE{\textbf{Plateau-Rayleigh instability of a soft layer coated on a rigid cylinder
}} \\
\vspace{0.3cm} & \vspace{0.3cm} \\

 & \noindent\large{Bharti, Andreas Carlson and Tak Shing Chan$^*$} \\
 
\includegraphics{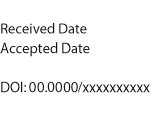} & \noindent\normalsize{
} We study the Plateau-Rayleigh instability of a viscoelastic soft solid layer coated on a rigid cylinder i.e.,  a soft fibre with a rigid core. The onset of instability is examined using a linear stability analysis.  We find that increasing the rigid cylinder radius reduce the growth rate of the fastest growing mode.  For each rigid cylinder radius,  a critical elastocapillary number is found below which all wavelengths of disturbances are stable. The critical value  for a soft fibre with a thick rigid cylindrical core can be  several orders of magnitudes larger than that for a totally soft fibre (no rigid core), which highlights  the strong stabilizing effect of the rigid core  on the system.  Increasing the relaxation timescale of the viscoelastic material also slows down the growth of disturbance, but has no effect on the critical elastocapillary number.  Interestingly, the wavelength  of the fastest growing mode is independent of the  rigid cylinder radius for the purely elastic case.  


\end{tabular}

 \end{@twocolumnfalse} \vspace{0.6cm}

  ]

\renewcommand*\rmdefault{bch}\normalfont\upshape
\rmfamily
\section*{}
\vspace{-1cm}


\footnotetext{\textit{$^{a}$~Mechanics Division, Department of Mathematics, University of Oslo, 0316 Oslo, Norway. Email: taksc@uio.no }}

\section{Introduction}
Significant stress is usually required for a solid to deform, and as such the effects due to surface tension have often been ignored.  Soft solids such as elastomers and gels \citep{Guo2020},  on the other hand,  have elastic moduli ranging  between kPa to MPa, which means they can deform much easier.   In recent years, there has been significant attention to how capillary effects can lead to soft solid deformations, and fascinating elastocapillary phenomena have been discovered \cite{Roman2010,Jagota2012,Wei2014,Style2017,Bico2018}.  The Plateau-Rayleigh instability (PRI)\cite{P43,R79a}, namely the instability driven by surface tension and often illustrated by the breaking up of liquid jets into droplets\cite{EV08},  has currently been examined for fibres made of soft solids \cite{Taffetani2015,Xuan2017,Mora2010,Lestringant2020,Fu2021,Tamim2021,Pandey2021,Dortdivanlioglu2022,Yang2022,Ru2022}.

 Studies of the PRI of soft fibres have focused on  soft materials that demonstrate elastic or viscoelastic responses \cite{Taffetani2015,Xuan2017,Mora2010,Lestringant2020,Fu2021,Tamim2021,Pandey2021,Dortdivanlioglu2022,Yang2022,Ru2022}.
 An experimental study using soft agar gel fibres by Mora et al.\cite{Mora2010} has shown that the instability occurs  when the elastocapillary length $\gamma/\mu \geq 6R$, where $R$ is the radius of the soft fibre, $\gamma$ is the solid surface tension and $\mu$ is the shear modulus.  Other studies of the PRI on a soft fibre include, for example, the formation of beads-on-string structures \cite{Bhat2010,Taffetani2015,Fu2021,Pandey2021}.  Another interesting factor that  might significantly modify the instability is having an inner rigid core, i.e.  a fibre consists of a soft-layer-coated on a  rigid cylinder.  Such kind of setup, but with a coated liquid film instead of a soft solid layer, has been studied extensively  \cite{Dav1999, Chang1999, KLIAKHANDLER2001,Ruy2008, Zheng2010,Haefner2015, Sad2017, Chen2018,Ji2019, Zhao2023} since  the early works  in the 1960s by Goren \cite{Goren1962, Goren1964}.  To name a few, some  studies address  the effects of  liquid slip on solid surface \cite{Haefner2015,Zhao2023} and  the dynamics of the droplets formed on the fibre \cite{KLIAKHANDLER2001}.  However,   there has been a lack of investigations on  situations in which the coated layer is a soft solid.  How viscoelastic properties and the rigid core influence the PRI of a soft-layer-coated fibre remain unclear, which is addressed in this article.
  
 Soft solids have nowadays been used in many applications,  for example in 3D bioprinting \cite{Zhang2021},  mimicking muscle tissues in biomedicine \cite{Fitzgerald2015} and water harvesting \citep{Nandakumar2019}.   In living organisms,  soft fibrous-shaped structures are often found in cellular tubes and compartments of cells.  Instability of these biological ingredients resembling to the PRI has been observed  \cite{Roy1994,Edouard2012,Gonzalez2015,King2020,Setru2021}.  One example is a recent study on the  undulation and droplet formation of a layer of condensed protein TPX2  on a microtubule \cite{Setru2021}.  Hence,  investigating  the PRI of fibrous  soft solids is becoming more important  for both understanding the  fundamental physical problems and the development of new technologies.   In this study, we examine the onset of PRI of a soft layer coated on a rigid cylinder through a linear stability analysis.
  
\section{Formulation}

\begin{figure}[!thbp]
	\begin{center}
		{\includegraphics[height=1.75 in]{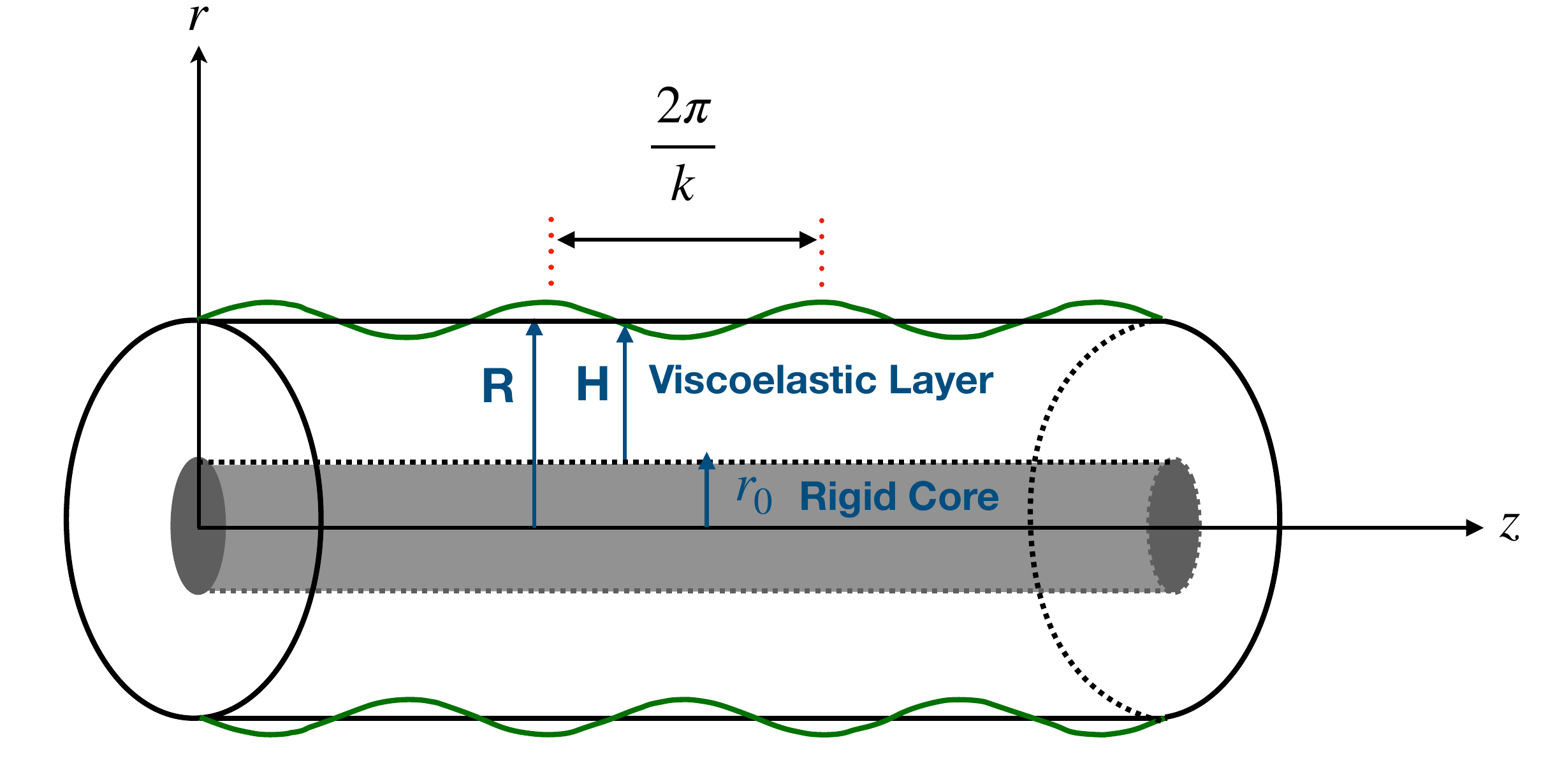}}
	\caption{Schematic representation of an infinitely long, rigid cylinder of radius $r_0$ coated with a layer of incompressible viscoelastic material of uniform thickness $H$ at an undeformed state.  A random disturbance of the interface of the viscoelastic layer is decomposed into  sinusoidal deformations of wavenumber $k$. In cylindrical coordinates systems, $r$ and $z$ are respectively the radial coordinate and the longitudinal coordinate.}
		\label{fig:1}
	\end{center}
\end{figure}

We consider a rigid cylinder of radius $r_0$ coated with a layer of incompressible viscoelastic material of uniform thickness $H$ at an undeformed state.   The whole fibre consisting of the rigid cylinder and the coated soft layer has  a total radius $R=r_0+H$ as shown in  Fig.\ref{fig:1}.  By neglecting all  body forces (e.g. gravity) and only considering  viscoelastic stresses, the equation of motion of a soft material element of density $\rho$ is given by
\begin{equation}\label{1}
\frac{\partial\sigma_{ij}(\bm{x},t)}{\partial x_j}=\rho\frac{\partial^{2}u_{i}(\bm{x},t)}{\partial t^{2}}
\end{equation}
where $\sigma_{ij}(\bm{x},t)$ and $u_{i}(\bm{x},t)$ are respectively the stress tensor and the displacement vector in index notation at a position vector $\bm{x}$ and time $t$.  Assuming a linear viscoelastic response of the incompressible soft material, the relation between $\sigma_{ij}$ and the strain tensor $\epsilon_{ij}$ is
\begin{equation}\label{2}
\sigma_{ij}(\bm{x},t)=2\int_{-\infty}^{t}\mu(t-t^{'})\frac{\partial\epsilon_{ij}(\bm{x},t^{'})}{\partial t^{'}}dt^{'}-p(\bm{x},t)\delta_{ij},
\end{equation}
where $\mu(t)$ is the shear relaxation function, $p$ is the pressure, $\delta_{ij}$ is the Kronecker delta, and the strain tensor $\epsilon_{ij}$ is related to the displacement as
\begin{equation}\label{3}
\epsilon_{ij}=\frac{1}{2}\Big(\frac{\partial u_{i}}{\partial x_{j}}+\frac{\partial u_{j}}{\partial x_{i}}\Big).
\end{equation}

There are different models that have been used to describe viscoelastic materials. The simplest models are  the Kelvin-Voigt model or the Maxwell model,  which consists of a spring and a viscous damper connected in parallel or in series respectively. 
In this study, we consider the soft material to behave as a gel described by the Chasset-Thirion model assuming a power law response given by
\begin{equation}\label{4}
\mu(t)=\mu_o\left[1+\Gamma(1-n)^{-1}\left(\frac{\tau}{t}\right)^n\right],
\end{equation}
where $\mu_{0}$ is the static shear modulus, $\tau$ is the relaxation timescale of the  viscoelastic response, $\Gamma$ is the gamma function and $n$ is a parameter typically smaller than or equal to unity.

To compute the growth rate of disturbance on the viscoelastic layer, we follow the approach delineated by previous studies \cite{Tamim2021,Zhao2023}. We decompose the time-dependent variables into normal modes $e^{st}$, where $s$ is the growth rate. The amplitude of the normal mode of a function $f(t)$ is obtained by the Laplace transform defined as    
\begin{equation}\label{5}
\tilde{f}(s)=\int_{0}^{\infty}f(t)e^{-st}\operatorname{d}t.
\end{equation}

We apply the Laplace transform to the constitutive relation (\ref{2}) and obtain
\begin{equation}\label{6}
\tilde{\sigma}_{ij}(\bm{x},s)=\hat{\mu}(s)\left [\frac{\partial \tilde{u}_{i}(\bm{x},s)}{\partial x_{j}}+\frac{\partial \tilde{u}_{j}(\bm{x},s)}{\partial x_{i}}\right ]-\tilde{p}(\bm{x},s)\delta_{ij}
\end{equation}
where $\tilde{\sigma}_{ij}$, $\tilde{u}_{i}$ and $\tilde{p}$ are respectively the Laplace transform of $\sigma_{ij}$, $u_{i}$ and $p$, and  $\hat{\mu}(s)$ is the shear modulus in Laplace space defined as 
\begin{equation}\label{7}
\hat{\mu}(s)\equiv s\int_{0}^{\infty}\mu(t)e^{-st}\operatorname{d}t=\mu_{0}[1+(s \tau)^{n}].
\end{equation}
Thus the governing equation (\ref{1}) in Laplace space can be written as
\begin{equation}\label{8}
\hat{\mu}(s)\frac{\partial^2 \tilde{u}_i}{\partial x_j\partial x_j}-\frac{\partial \tilde{p}}{\partial x_i}=\rho s^{2}\tilde{u}_i.
\end{equation}
Next, we scale the lengths with the radius of the whole fibre $R$, the time with the capillary timescale $\sqrt{\rho R^{3}/\gamma}$ and the stresses by $\mu_o$. We define the following dimensionless variables as 
\begin{eqnarray}
\bar{r}=\frac{r}{R}, \
\bar{z}=\frac{z}{R}, \
\bar{s}=\sqrt{\frac{\rho R^{3}}{\gamma}}s, \\
\bar{u}_i(\bar{r}, \bar{z}, \bar{s})=\frac{\tilde{u}_i}{R}, \
\bar{p}(\bar{r}, \bar{z}, \bar{s})=\frac{\tilde{p}}{\mu_{0}}, \
\bar{\sigma}_{ij}(\bar{r}, \bar{z}, \bar{s})=\frac{\tilde{\sigma}_{ij}}{\mu_{0}}. 
\end{eqnarray}

The dimensionless form of the governing equation (eq. \ref{8}) in Laplace space is
\begin{equation}\label{9}
\Big[1+(\bar{\tau}\bar{s})^{n}\Big]\frac{\partial^2 \bar{u}_i}{\partial x_j\partial x_j}-\frac{\partial \bar{p}}{\partial x_i}=\Sigma  \bar{s}^{2}\bar{u}_i
\end{equation}
where $\bar{\tau}=\tau/\sqrt{\rho R^{3}/\gamma}$ and the elastocapillary number $\Sigma=\gamma/\left(\mu_{0}R\right)$.

We consider only the longitudinal disturbance and neglect the azimuthal disturbance as azimuthal normal modes always increase the surface energy \cite{EV08}. The deformation of the soft layer is axisymmetric. We hence use the cylindrical coordinate system $(r, \phi,z)$ and the corresponding unit vectors are denoted as ($\hat{\bm{r}},\hat{\bm{\phi}},\hat{\bm{z}}$). The displacement vector is denoted as $\bm{u}(r,z,t)=u_r(r,z,t)\hat{\bm{r}}+ u_z(r,z,t)\hat{\bm{z}}$.   Note that $u_{\phi}=0$ due to axisymmetry.

The governing equation (\ref{9}) can be solved by applying the Helmholtz
decomposition of the displacement  as described in the references \cite{Tamim2021,Zhao2023}. 
The general solutions are given as 
\begin{equation}\label{10}
\bar{p}(\bar{s})=-\bar{s}^{2}\Sigma\Big[A_{1}I_{0}(\bar{k}\bar{r})+A_{3}K_{0}(\bar{k}\bar{r})\Big]e^{i\bar{k}\bar{z}},
\end{equation}
\begin{equation}\label{11}
\bar{u}_{r}=\Big[A_{1}\bar{k}I_{1}(\bar{k}\bar{r})-A_{2}(i\bar{k})I_{1}(\alpha \bar{r})-A_{3}\bar{k}K_{1}(\bar{k}\bar{r})-A_{4}(i\bar{k})K_{1}(\alpha\bar{r})\Big]e^{i\bar{k}\bar{z}},
\end{equation}

and
\begin{equation}\label{12}
\bar{u}_{z}=\Big[A_{1}(i\bar{k})I_{0}(\bar{k}\bar{r})+A_{2}\alpha I_{0}(\alpha\bar{r})+A_{3}(i\bar{k})K_{0}(\bar{k}\bar{r}) -A_{4}\alpha K_{0}(\alpha\bar{r})\Big]e^{i\bar{k}\bar{z}}
\end{equation}
where $\bar{k}\equiv kR$ is the dimensionless wavenumber, $I$ and $K$  are respectively the modified Bessel functions of the first and second kind, $\alpha=\sqrt{\bar{s}^{2}\beta+\bar{k}^{2}}$,  $\beta=\Sigma/(1+(\bar{\tau}\bar{s})^{n})$ and  $A_{m}(m=1,2,3,4)$ are the undermined coefficients.

Next we impose the boundary conditions. At  $\bar{r}=\bar{r}_{0}\equiv r_0/R$, there is no penetration of material, thus
  \begin{equation}\label{13}
 \bar{u}_{r}\vert_{\bar{r}=\bar{r}_{0}}=0.
 \end{equation}
 For the z-direction, we impose a no-displacement condition
  \begin{equation}\label{14}
\bar{u}_{z}\vert_{\bar{r}=\bar{r}_{0}}=0.
 \end{equation}

At $\bar{r}=1$, we assume the slope of the deformed interface to be small. Balancing the Laplace pressure due to solid surface tension and the viscoelastic stresses gives \begin{equation}\label{15}
\bar{\sigma}_{rr}\vert_{\bar{r}=1}= \Sigma\left(\frac{\partial^{2}\bar{u}_{r}}{\partial \bar{z}^{2}}+\bar{u}_{r}\right)\vert_{\bar{r}=1},
\end{equation} 
 in r-direction, and gives the vanishing shear stress
 \begin{equation}\label{16}
\bar{ \sigma}_{rz}\vert_{\bar{r}=1}=0
 \end{equation}
  in z-direction.
 
\section{Dispersion relation}
Using the expressions of the general solutions (eq. \ref{10}-\ref{12}) for the boundary conditions  (eq. \ref{13}-\ref{16})  yields the following set of linear equations for the unknowns $A_m$.  

\begin{equation}\label{17}
A_{1}\bar{k}I_{1}(\bar{r}_{0}\bar{k})-A_{2}i\bar{k}I_{1}(\bar{r}_{0}\alpha)-A_{3}\bar{k}K_{1}(\bar{r}_{0}\bar{k})-A_{4}i\bar{k}K_{1}(\bar{r}_{0}\alpha) =0,
\end{equation}

\begin{equation}\label{18}
 A_{1}(i\bar{k})I_0(\bar{r}_{0}\bar{k})+A_2\alpha I_0(\bar{r}_{0}\alpha)+A_3(i\bar{k})K_0(\bar{r}_{0}\bar{k})-A_4\alpha K_0(\bar{r}_{0}\alpha)=0,
\end{equation}

\begin{equation}\label{19}
\begin{split}
& A_{1}\Bigl\{\frac{\alpha^{2}+\bar{k}^{2}}{2}I_{0}(\bar{k})-\bar{k}I_{1}(\bar{k})-\frac{\beta}{2}(1-\bar{k}^{2})\bar{k}I_{1}(\bar{k})\Bigr\}\\
& + A_{2}\Bigl\{-i\bar{k}\Big[\alpha I_{0}(\alpha)-I_{1}(\alpha)\Big] +\frac{\beta}{2}i\bar{k}(1-\bar{k}^{2})I_{1}(\alpha)\Bigr\} \\ 
& +A_{3}\Bigl\{\frac{\alpha^{2}+\bar{k}^{2}}{2}K_{0}(\bar{k})+\bar{k}K_{1}(\bar{k})+\frac{\beta}{2}(1-\bar{k}^{2})\bar{k}K_{1}(\bar{k})\Bigr\}\\
& +A_{4}\Big\{i\bar{k}\Big[\alpha K_{0}(\alpha)+K_{1}(\alpha)\Big]+\frac{\beta}{2}i\bar{k}(1-\bar{k}^{2})K_{1}(\alpha)\Bigr\}
=0,
\end{split}
\end{equation}
and
\begin{equation}\label{20}
\begin{split}
& A_{1}2i\bar{k}^{2}I_{1}(\bar{k})+A_{2}(\bar{k}^{2}+\alpha^{2})I_{1}(\alpha)-A_{3}2i\bar{k}^{2}K_{1}(\bar{k})\\
& +A_{4}(\bar{k}^{2}+\alpha^{2})K_{1}(\alpha)=0.
\end{split}
\end{equation}

The solvability condition for non-trivial solutions of the linear equations with unknowns $A_m$ is that the determinant is vanishing, which gives the following dispersion relation \\
\begin{equation}\label{disp}
 \begin{vmatrix}
 \bar{k}I_{1}(\bar{r}_{0}\bar{k})& -i\bar{k}I_{1}(\bar{r}_{0}\alpha)  & -\bar{k}K_{1}(\bar{r}_{0}\bar{k}) &  -i\bar{k}K_{1}(\bar{r}_{0}\alpha)\\
  (i\bar{k})I_0(\bar{r}_{0}\bar{k}) &  \alpha I_0(\bar{r}_{0}\alpha)& (i\bar{k})K_0(\bar{r}_{0}\bar{k})& -\alpha K_0(\bar{r}_{0}\alpha)\\ 
  C_{1} &C_{2} & C_{3} & C_{4} \\
    2i\bar{k}^{2}I_{1}(\bar{k})  & (\bar{k}^{2}+\alpha^{2})I_{1}(\alpha) & -2i\bar{k}^{2}K_{1}(\bar{k}) & (\bar{k}^{2}+\alpha^{2})K_{1}(\alpha)
\end{vmatrix}=0,
\end{equation}
where \\
$C_{1}= \frac{\alpha^{2}+\bar{k}^{2}}{2}I_{0}(\bar{k})-\bar{k}I_{1}(\bar{k})-\frac{\beta}{2}(1-\bar{k}^{2})\bar{k}I_{1}(\bar{k})$,\\
$C_{2}=-i\bar{k}\Big(\alpha I_{0}(\alpha)-I_{1}(\alpha)\Big) +\frac{\beta}{2}i\bar{k}(1-\bar{k}^{2})I_{1}(\alpha)$,\\
$C_{3}=\frac{\alpha^{2}+\bar{k}^{2}}{2}K_{0}(\bar{k})+\bar{k}K_{1}(\bar{k})+\frac{\beta}{2}(1-\bar{k}^{2})\bar{k}K_{1}(\bar{k})$,\\
$C_{4}=i\bar{k}\Big(\alpha K_{0}(\alpha)+K_{1}(\alpha))+\frac{\beta}{2}i\bar{k}(1-\bar{k}^{2})K_{1}(\alpha)$.\\
The dimensionless control parameters are: $\bar{r}_{0}$, $\Sigma$, $\bar{\tau}$ and $n$.

\section{Results}
The validation of the dispersion relation (eq. \ref{disp}) is presented in the Appendix in which we compare our results in the Newtonian fluid limit with some previous studies. In this section, we first consider the purely elastic solid limit and then we focus on the effects of viscoelasticity.

\subsection{The purely elastic solid limit}
When the viscoelastic relaxation timescale $\tau$ is small compared to the time scale $\sqrt{\rho R^{3}/\gamma}$, i.e. taking the limit $\bar{\tau} \rightarrow 0$, our viscoelastic model reduces to the purely elastic model. In this limit, $\alpha=\sqrt{\bar{s}^{2}\Sigma+\bar{k}^{2}}$ and $\beta = \Sigma$ in the dispersion relation (eq. \ref{disp}).  We examine how the instability depends on the dimensionless parameters $\Sigma$ and $\bar{r}_0$.
\subsubsection{Dependence on the rigid core radius and the elastocapilliary number} \label{rigidCore}
\begin{figure}[!thbp]
\begin{center}
\includegraphics[height=1.9 in]{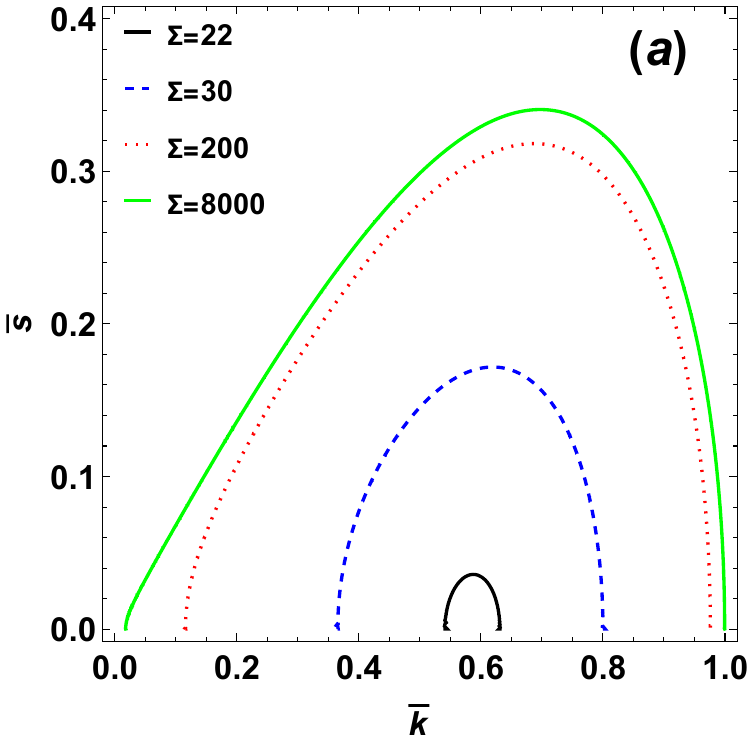}
\includegraphics[height=1.6 in]{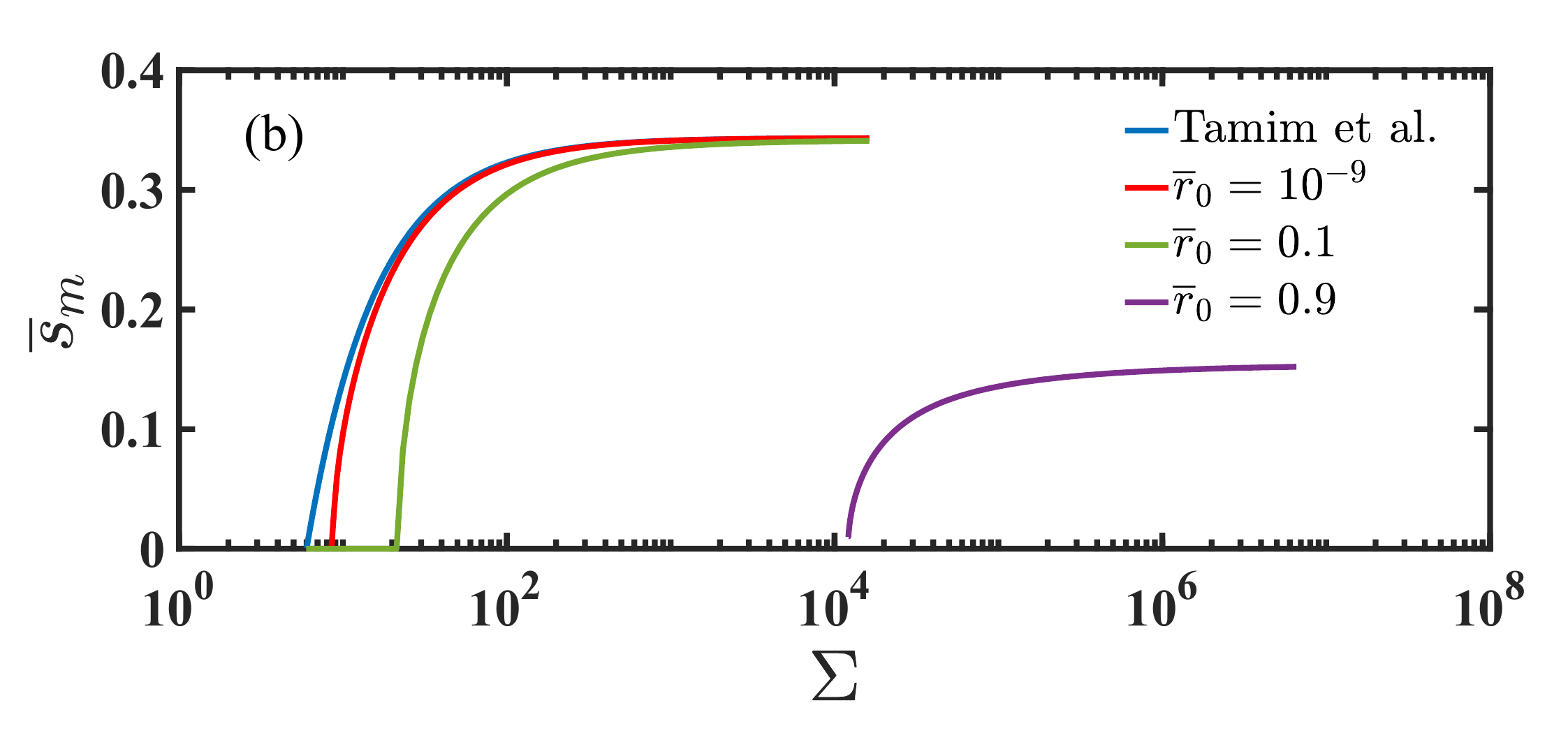}
\includegraphics[height=1.55 in]{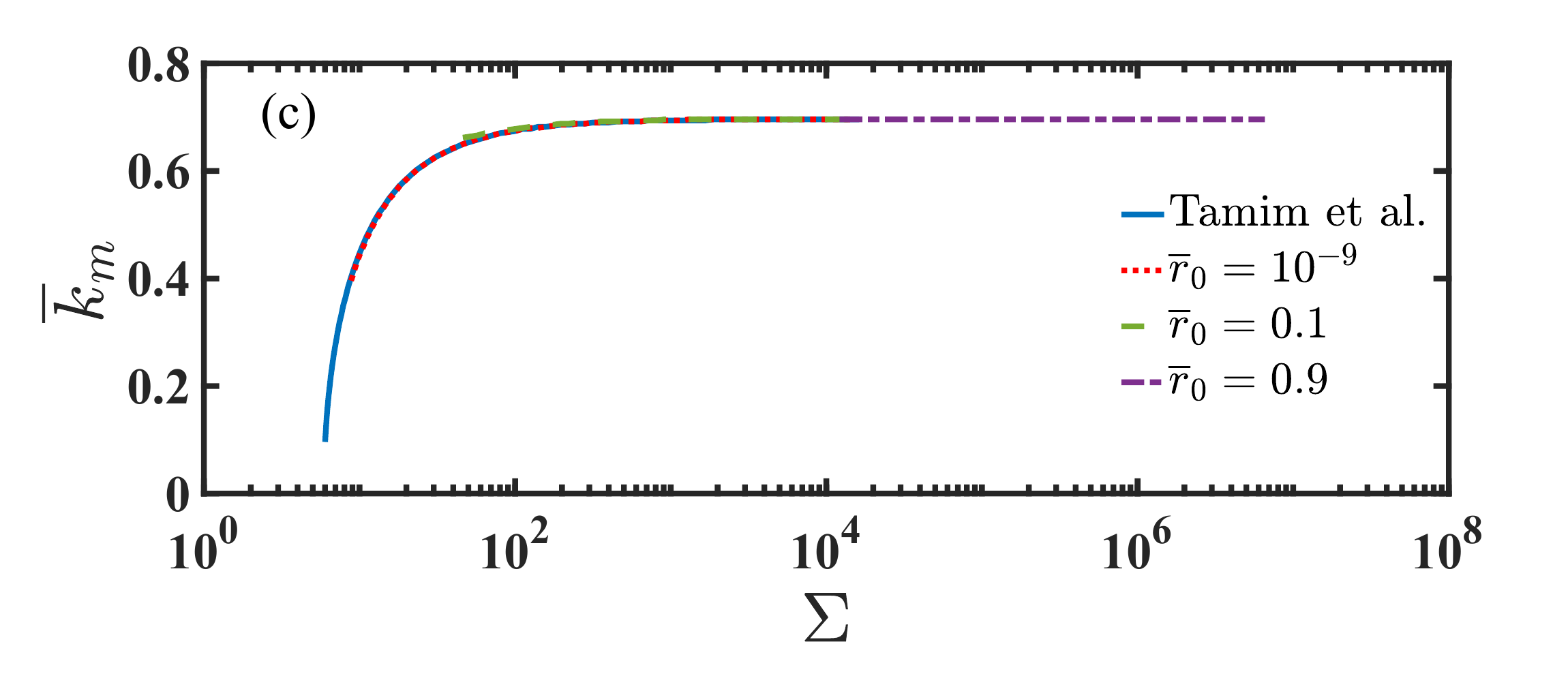}
\includegraphics[height=1.5 in]{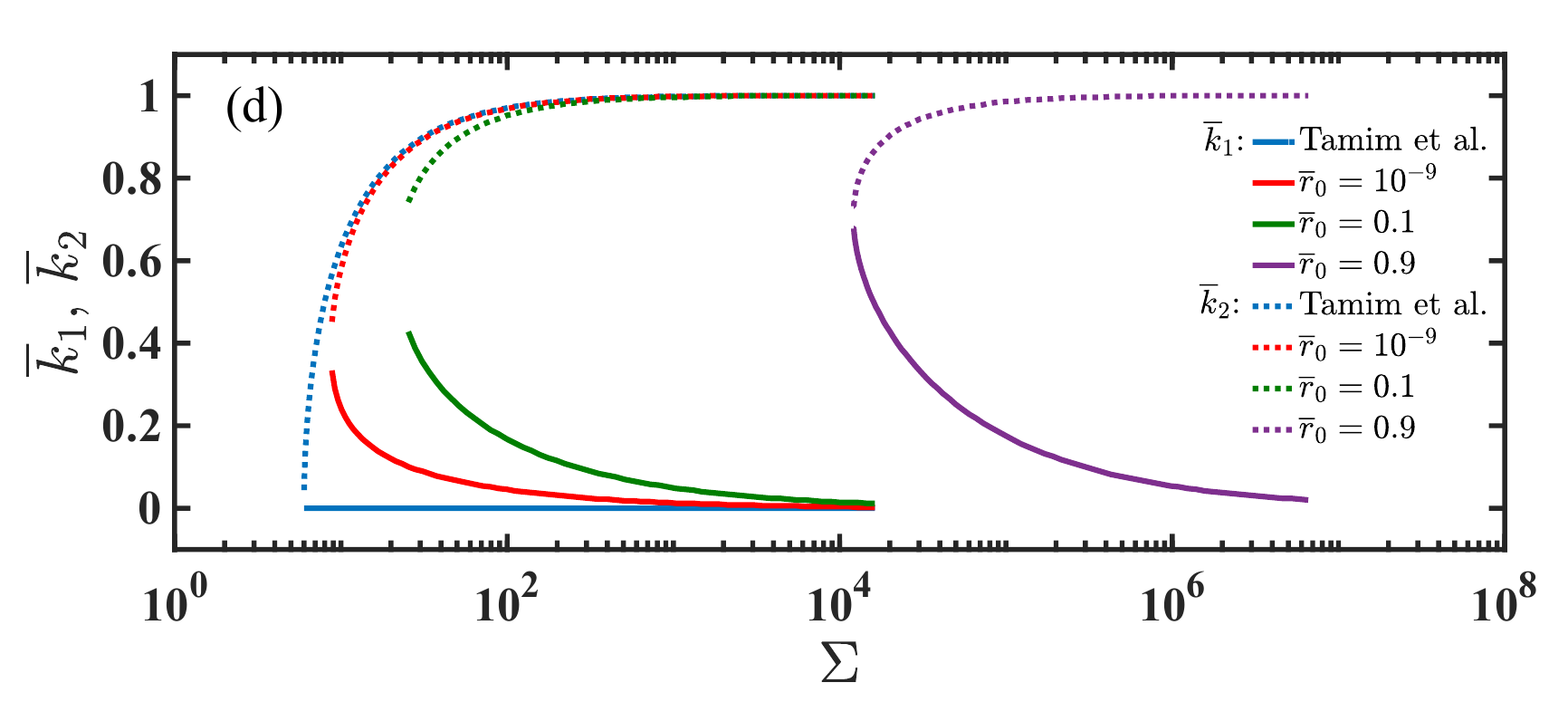}
\caption{(a) The dimensionless growth rate $\bar{s}$ as a function of the dimensionless wavenumber $\bar{k}$ for different values of $\Sigma$ and a fixed $\bar{r}_{0}=0.1$ and $\bar{\tau} =0$ (purely elastic).  (b), (c) and (d): The characteristic quantities  $\bar{s}_{m}$ in (b),  $\bar{k}_{m}$ in (c), and  $\bar{k}_{1}$, $\bar{k}_{2}$ in (d) as a function of $\Sigma$  for four different dimensionless rigid core radius, i.e. $\bar{r}_0=0$ (from Tamim \emph{et al.} \cite{Tamim2021}), $\bar{r}_0=10^{-9}$, $\bar{r}_0=0.1$  and $0.9$.  Other parameters: $\bar{\tau} =0$.}
\label{fig:2}
\end{center}
\end{figure}
We show the dispersion relation (eq.  \ref{disp}) by plotting the dimensionless growth rate $\bar{s}$ as a function of the dimensionless wavenumber $\bar{k}$  for different values of $\Sigma$ and a fixed $\bar{r}_0=0.1$ in Fig. \ref{fig:2} (a).  We see that for each curve, the unstable modes (i.e. $\bar{s}>0$) lie within a certain range of dimensionless wavenumber,  i.e. $\bar{k}_1<\bar{k}<\bar{k}_2$, where $\bar{k}_1$ and $\bar{k}_2$ are defined as $\bar{s}(\bar{k}_1)=0$ and $\bar{s}(\bar{k}_2)=0$.  Within each range of unstable modes, there is a dimensionless wavenumber $\bar{k}=\bar{k}_{m}$ that corresponds to the fastest growing mode with a   maximum dimensionless growth rate $\bar{s}_{m}\equiv\bar{s}(\bar{k}=\bar{k}_{m})$.   
To study how the  characteristic quantities depend on the control parameters,  we plot $\bar{s}_{m}$, $\bar{k}_{m}$, and $\bar{k}_1$(and $\bar{k}_2$) as a function of $\Sigma$  respectively  in Fig. \ref{fig:2} (b), (c) and (d) for three different dimensionless rigid core radius, i.e. $\bar{r}_0=10^{-9}$, $\bar{r}_0=0.1$  and $0.9$. We also add the result from Tamim \emph{et al.} \cite{Tamim2021} for the situation of a soft fibre without a rigid cylindrical core (i.e.   $\bar{r}_0=0$).  We  see in Fig. \ref{fig:2} (b) that for all four different $\bar{r}_0$,  the dimensionless growth rate of the fastest growing mode $\bar{s}_{m}$ decreases with decreasing  $\Sigma$.  When $\Sigma$ is reduced to a critical value $\Sigma_{c}$,   $\bar{s}_{m}$ drops to zero. There is no positive solution of $\bar{s}$ for  $\Sigma<\Sigma_c$.  It means the coated elastic layer is stable under disturbance of any wavelength when $\Sigma<\Sigma_c$.  We would also point out that, first, the critical value $\Sigma_{c}$ for $\bar{r}_0=0.9$   is orders of magnitude larger than that for $\bar{r}_0=0.1$.  Second,  even for the rigid cylindrical core radius as small as $\bar{r}_0=10^{-9}$, there is a slight difference from the result for $\bar{r}_0=0$ when $\Sigma$ is close to $\Sigma_c$.   A stability phase diagram of $\Sigma_c$ vs $\bar{r}_0$ will be examined in section \ref{spd}.  
Interestingly,  as shown in  Fig. \ref{fig:2} (c),  we  find that the dimensionless wavenumber of the fastest growing mode $\bar{k}_{m}$ is independent of the dimensionless radius of the rigid core  $\bar{r}_0$,  and decreases with  reducing $\Sigma$.  When $\Sigma\rightarrow\infty$,  we find $\bar{k}_{m}=0.7$,  which agrees with the dimensionless wavenumber of the fastest growing mode  for the classical PRI of  inviscid fluid jet \citep{R79a,EV08}.  Another remarkable point is that  $\bar{k}_{m}$ starts to drop significantly with reducing $\Sigma$ only when $\Sigma\lesssim 10^2$.  For cases with $\Sigma_c>10^2$, e.g. for  $\bar{r}_0=0.9$,   the dimensionless wavenumber of the fastest growing mode is always close to the asymptotic value, i.e. $\bar{k}_{m}\approx 0.68$.  Regarding the range of  unstable modes,  as we can see in Fig. \ref{fig:2} (a) and Fig. \ref{fig:2} (d), it shrinks when $\Sigma$ is reduced.  Namely,  $\bar{k}_1$ increases (or remains zero for $\bar{r}_0=0$)  and $\bar{k}_2$ decreases. 

\subsubsection{Stability phase diagram}\label{spd}
\begin{figure}[!thbp]
\begin{center}
\includegraphics[height=1.35 in]{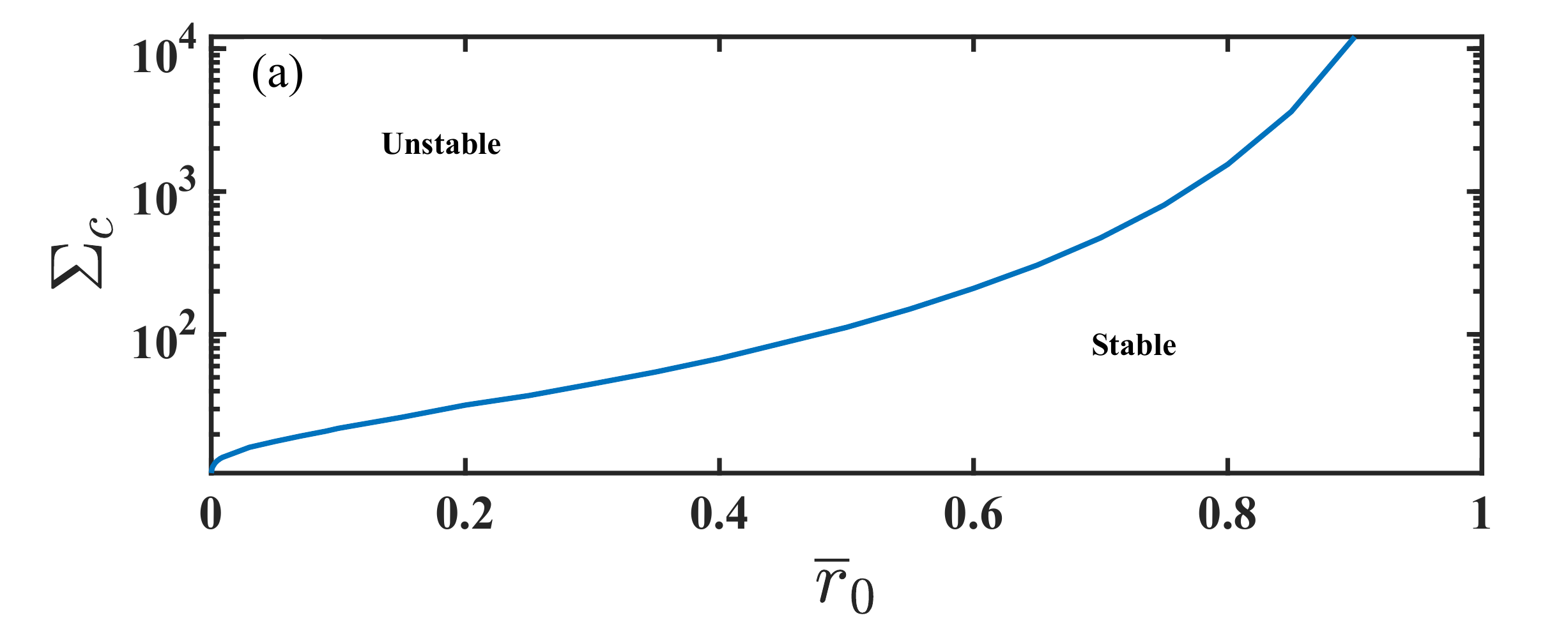}
\includegraphics[height=1.05 in]{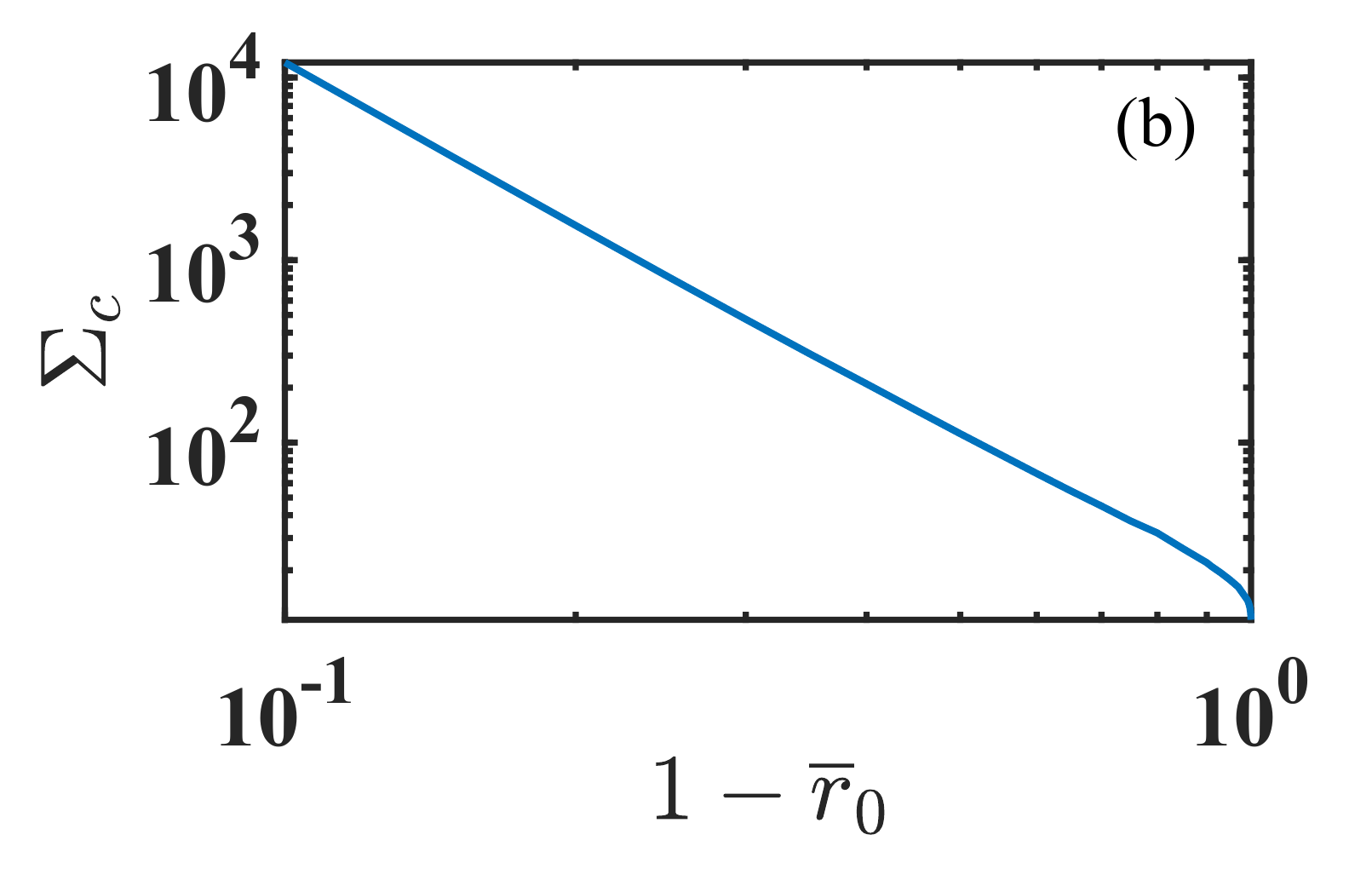}
\includegraphics[height=1.05 in]{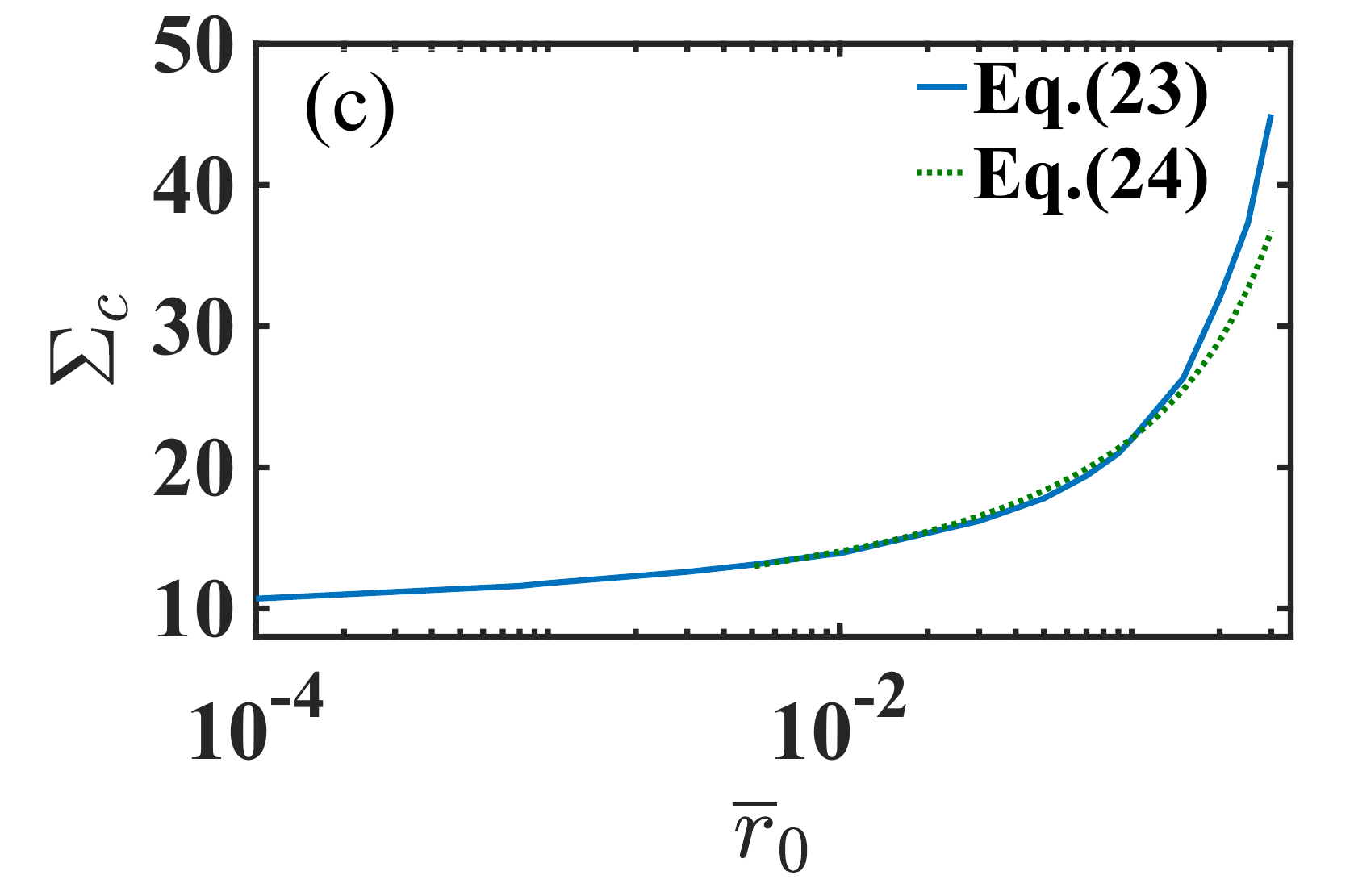}
\caption{Stability phase diagram for the purely elastic case,  i.e. $\bar{\tau}=0$. (a) The critical  elastocapillary number $\Sigma_c$ (log scale) as a function of $\bar{r}_{0}$.  (b) The critical  elastocapillary number $\Sigma_c$ as a function of $1-\bar{r}_{0}$ in log-log scales. (c) A linear-log plot of the critical  elastocapillary number $\Sigma_c$ as a function of $\bar{r}_{0}$.}
\label{fig:3}
\end{center}
\end{figure}
To show how the onset of instability depends on $\Sigma$ and $\bar{r}_0$, we plot a phase diagram of $\Sigma_{c}$ (in log scales) as a function of $\bar{r}_0$ in Fig. \ref{fig:3} (a).  We  see that increasing the radius of the rigid cylindrical core or decreasing the elastocapillary number can make the coated film more stable.  Hence both the stiffness and the rigid core perform a stabilizing effect.  In  Fig. \ref{fig:3}(b), we plot $\Sigma_{c}$ as a function of the thickness of the coated layer, i.e.  $\bar{H}=1-\bar{r}_{0}$, in log-log scales.  We find that  when $\bar{r}_{0} \gtrsim 0.2$,   the curve follows a power law $\Sigma_{c}\sim \bar{H}^{\lambda}$ where $\lambda \approx -2.75$.  Hence the critical value $\Sigma_{c}$ for a thin coated elastic film is orders of magnitudes larger than a thick film.  In the opposite limit of $\bar{r}_{0}\rightarrow 0$,  we find that our results can be described by a logarithmic relation given as 
\begin{equation}\label{21}
\Sigma_{c}=\Sigma_{co}-\frac{a}{\log(\bar{r}_{0})}
\end{equation}
where $\Sigma_{co}=6$ is the critical value for a soft elastic fibre without a rigid cylindrical core obtained by Mora et al.\cite{Mora2010}  or Tamim \emph{et al.}  \citep{Tamim2021},  and $a=37$ is a fitting parameter.  The comparison is shown in Fig. \ref{fig:3} (c) for small $\bar{r}_{0}$.
\subsection{Viscoelastic effects}
\begin{figure}[!thbp]
\begin{center}
\includegraphics[height=2.25 in]{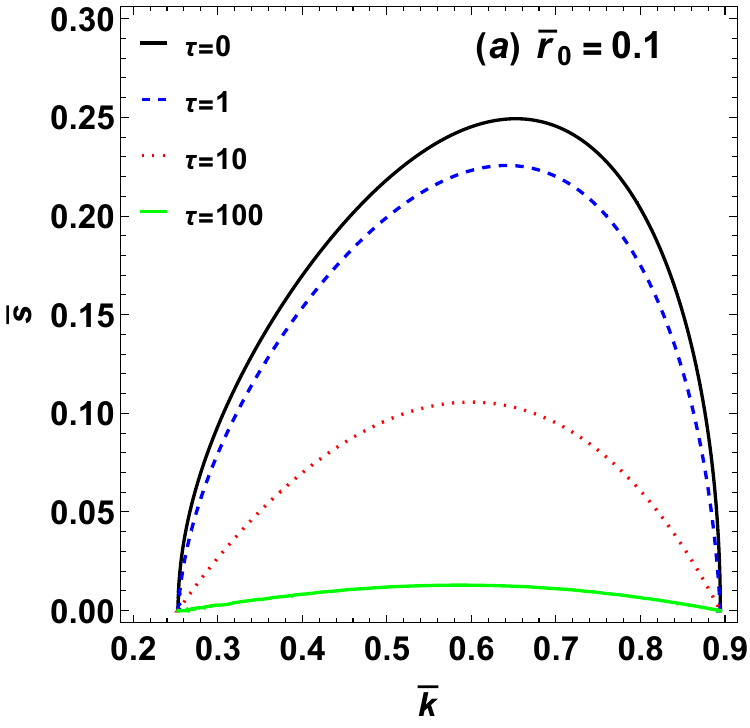}
\includegraphics[height=2.25 in]{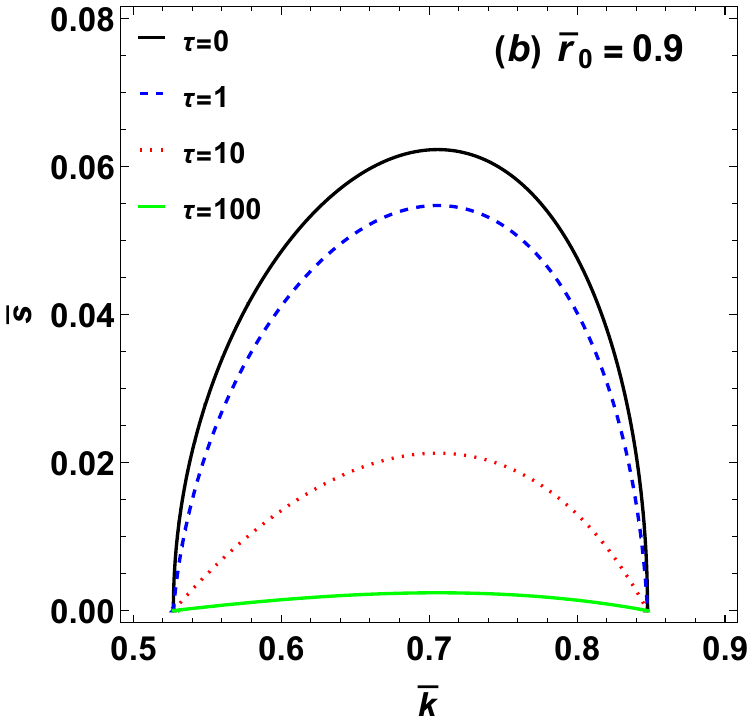}
\includegraphics[height=2.25 in]{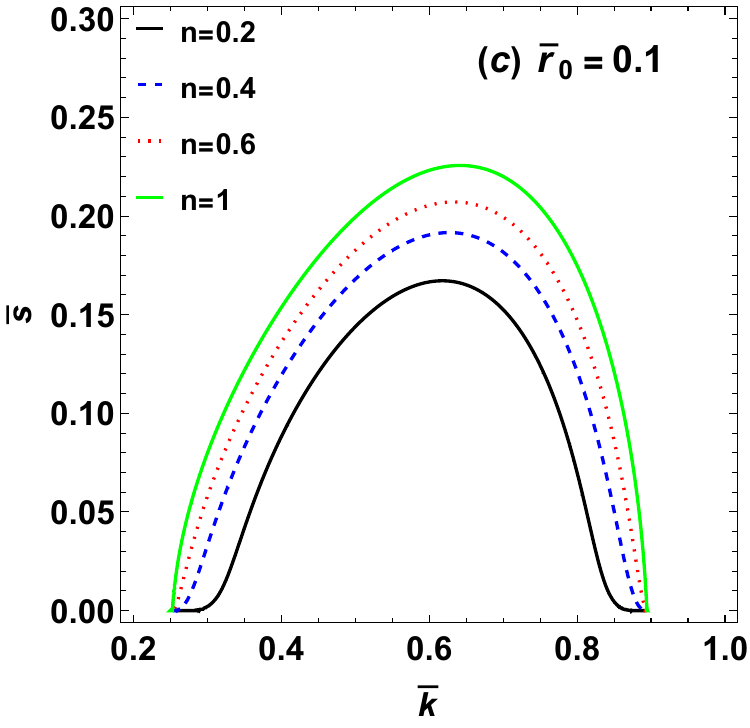}
\includegraphics[height=2.25 in]{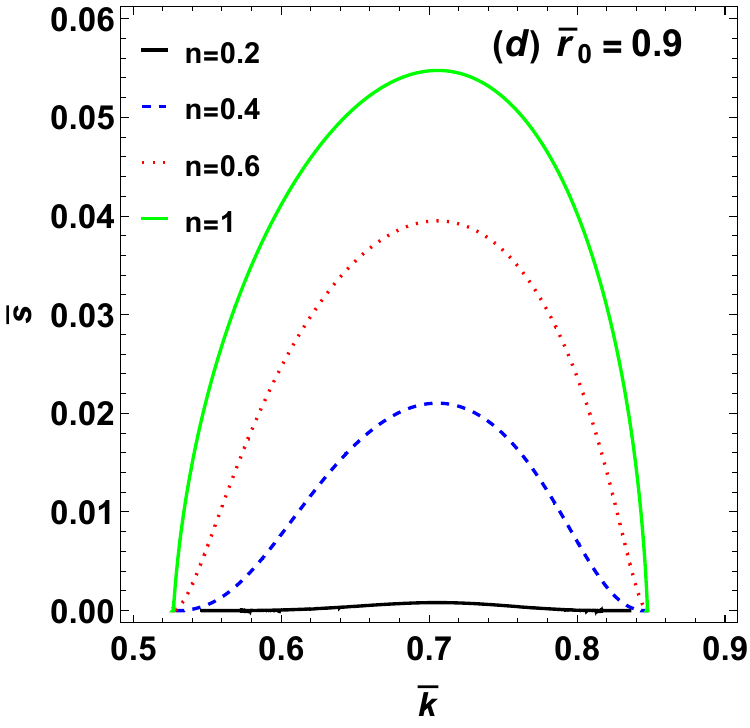}
\caption{The dimensionless growth rate $\bar{s}$ as a function of the dimensionless wavenumber $\bar{k}$ for different values of $\bar{\tau}$ and a fixed value of exponent $n=1$  with  $\bar{r}_{0}=0.1$ and $\Sigma=50$ in (a),  and  $\bar{r}_{0}=0.9$ and $\Sigma=15000$ in (b).  The dimensionless growth rate $\bar{s}$ vs the dimensionless wavenumber $\bar{k}$ for different values of exponent $n$ and a fixed value of $\bar{\tau}=1$  with  $\bar{r}_{0}=0.1$ and $\Sigma=50$ in (c),  and  $\bar{r}_{0}=0.9$ and $\Sigma=15000$ in (d).  }
\label{fig:4}
\end{center}
\end{figure} 

The dimensionless growth rate $\bar{s}$ as a function of the dimensionless wavenumber $\bar{k}$ for different values of $\Sigma$ and a fixed $\bar{r}_{0}=0.1$ and $\bar{\tau} =0$ (purely elastic).  (b), (c) and (d):

\begin{figure}[!thbp]
\begin{center}
\includegraphics[height=1.5 in]{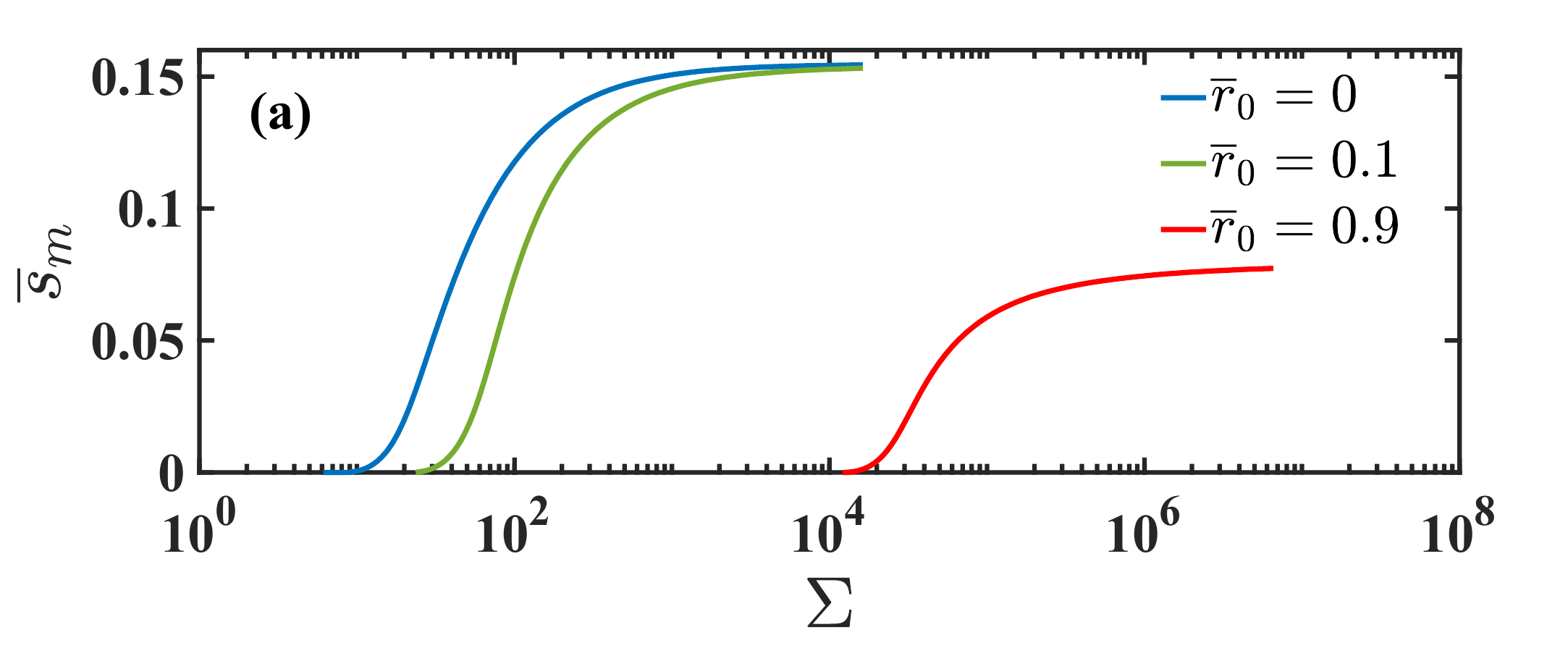}
\includegraphics[height=1.25 in]{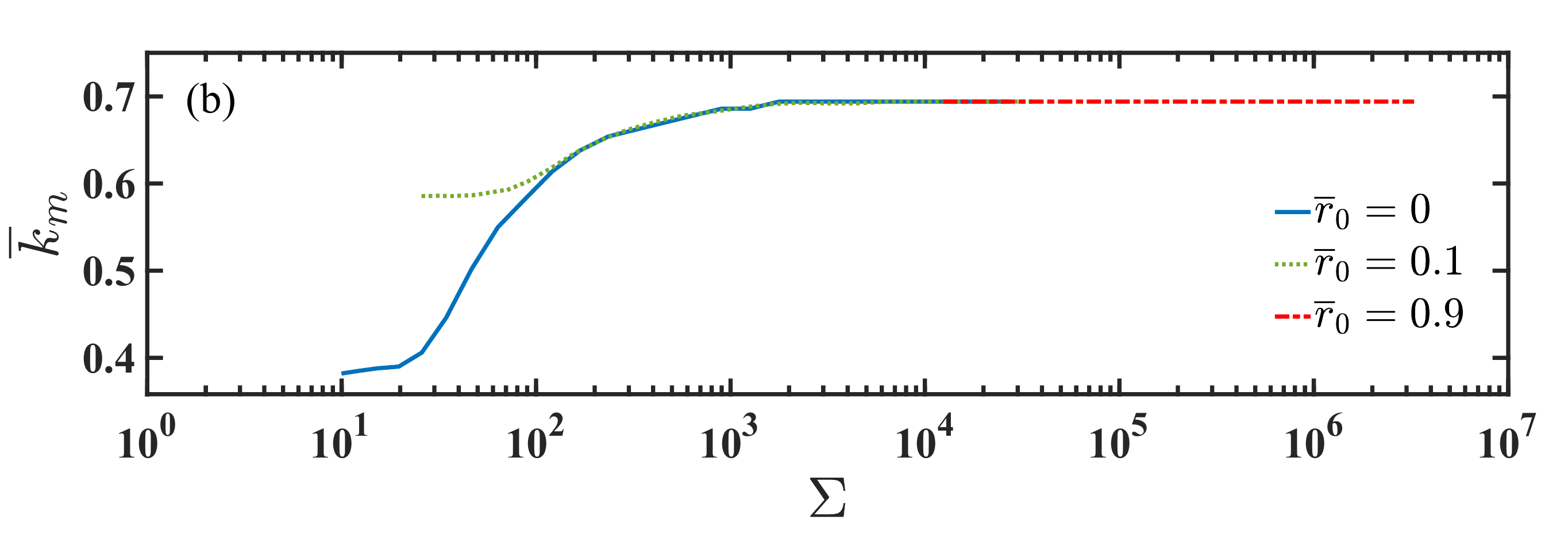}
\caption{ The characteristic quantities  $\bar{s}_{m}$ in (a) and $\bar{k}_{m}$ in (b) as a function of $\Sigma$  for three different dimensionless rigid core radius, i.e. $\bar{r}_0=0$ (from Tamim \emph{et al.} \cite{Tamim2021}), $\bar{r}_0=0.1$  and $0.9$.  Other parameters: $\bar{\tau}=100$ and $n=0.5$.}
\label{fig:5}
\end{center}
\end{figure}

We study the instability when $\bar{\tau}\neq 0$. We show  results for  two different rigid cylindrical core radius, namely $\bar{r}_{0}=0.1$ (a thick coated layer)  and $\bar{r}_{0}=0.9$ (a thin coated layer).  Since the critical elastocapillary numbers for $\bar{r}_{0}=0.1$ and $\bar{r}_{0}=0.9$ are different by orders of magnitude,  we take a fixed value of  $\Sigma=50$ for the case of $\bar{r}_{0}=0.1$ and a fixed value of  $\Sigma=15000$ for the case of $\bar{r}_{0}=0.9$.  In  Fig.\ref{fig:4} (a) and (b), we present the dispersion relation from eqn. (\ref{disp}) for different  $\bar{\tau}$ and  in Fig.\ref{fig:4} (c) and (d) for different $n$.  We see that both $\bar{k}_1$ and $\bar{k}_2$ are independent of $\bar{\tau}$ and $n$.  The dimensionless growth rate of  the fastest growing mode $\bar{s}_m$ decreases when $\bar{\tau}$ is enhanced.  It means that the viscoelastic relaxation of the material slows down the growth of disturbance.  We also see that when $\bar{\tau}$ is increased,  the dimensionless wavenumber of the fastest growing mode becomes smaller  for $\bar{r}_{0}=0.1$ but remains constant for $\bar{r}_{0}=0.9$.  When varying the other viscoelastic parameter $n$,  the dimensionless growth rate  of the fastest growing mode $\bar{s}_m$  decreases with decreasing $\bar{n}$.  The change  of $\bar{s}_m$ with varying  $\bar{n}$ is more sensitive for $\bar{r}_{0}=0.9$ than for $\bar{r}_{0}=0.1$ as we can see in Fig.\ref{fig:4} (c) and (d).  The dimensionless wavenumber of  the fastest growing mode $\bar{k}_m$  decreases slightly when $n$ is increased from $n=0.2$ to $n=1$ for $\bar{r}_{0}=0.1$ but remains constant  for $\bar{r}_{0}=0.9$. 
Lastly, we take $\bar{\tau}=100$ and $n=0.5$, and  show the results for $\bar{s}_m$ and $\bar{k}_m$  as a function of $\Sigma$ for three different $\bar{r}_{0}=0$,  $0.1$ and $0.9$ in Fig.\ref{fig:5}. We see the behaviors of $\bar{s}_m$ and $\bar{k}_m$ for $\bar{\tau}=100$ are similar to those shown in Fig.\ref{fig:2} (b) and (c) for the purely elastic caes, i.e.$\bar{\tau}=0$, except when $\Sigma$ is close to the critical value $\Sigma_c$.  For $\bar{\tau}=100$,  the critical values $\Sigma_c$ are the same as for $\bar{\tau}=0$. However, when  $\Sigma$ is approaching $\Sigma_c$,  the curves for both $\bar{s}_m$ and $\bar{k}_m$ bend concavely.  Moreover, the value of $\bar{k}_m$ becomes dependent on $\bar{r}_{0}$ when $\Sigma$ is around $\Sigma_c$.

\section{Conclusion}
 We investigate the onset of the PRI of a soft layer coated on a rigid  cylinder by analytically deriving the dispersion relation using the linear stability analysis.  We implement the Chasset-Thirion model for the viscoelastic response of the soft layer.  We find that the stiffness (characterized by $1/\Sigma$) and the rigid cylindrical core  (characterized by $\bar{r}_{0}$) perform a stabilizing effect.  The dimensionless growth rate  of the fastest growing mode $\bar{s}_m$ decreases with decreasing  $\Sigma$ or increasing $\bar{r}_{0}$.  Importantly, there exists a critical elastocapillary number $\Sigma_c$  for each $\bar{r}_{0}$ such that the coated layer is stable for any $\Sigma<\Sigma_c$.  The critical value depends strongly on $\bar{r}_{0}$.  For example, for $\bar{r}_{0}=0.9$, the soft layer becomes unstable only when the fibre is very soft, namely when  $\Sigma\geq 12100 $. While for a soft fibre without a rigid core ($\bar{r}_{0}=0$),  the onset of instability occurs at $\Sigma \approx 6 $, which is three to four orders of  magnitudes smaller.  This remarkable result would be interesting to be verified by performing experiments for different soft coatings on a fibre.   Regarding the dimensionless wavenumber  of the fastest growing mode $\bar{k}_m$,  we find that $\bar{k}_m$ decreases with decreasing  $\Sigma$.  Interestingly,  $\bar{k}_m$ is independent of $\bar{r}_{0}$ for the purely elastic case.
  
  Regarding the roles of  the viscoelastic parameters $\bar{\tau}$  and $n$,  we find that increasing  the relaxation timescale of viscoelastic material $\bar{\tau}$ or reducing the power  $n$ can  slow down the growth of disturbance.  Changing either $\bar{\tau}$  or $n$ have no effect on the critical elastocapillary number.  However,  for $\bar{\tau}\neq 0$,  the curves for both $\bar{s}_m$ and $\bar{k}_m$ bend concavely when  $\Sigma$ is approaching $\Sigma_c$.  The value of $\bar{k}_m$ becomes dependent on $\bar{r}_{0}$ when $\Sigma$ is around $\Sigma_c$.
 
\section{Appendix: validation of our model}

\subsection{The Newtonian fluid limit}
Our viscoelastic model reduces to the Newtonian fluid when $n=1$ and in the limits of $\mu_o\rightarrow 0$ and $\mu_o\tau\rightarrow \eta$, where $\eta$ is the dynamic viscosity of the fluid. In terms of the dimensionless parameters, it means $\Sigma\rightarrow \infty$ and $\Sigma/\tilde{\tau}\rightarrow \operatorname{Oh}\equiv \eta/\sqrt{\rho \gamma R}$, where $\operatorname{Oh}$ is called the Ohnesorge number.  In this limiting situation, the dimensionless governing equation (\ref{9}) is reduced to 
\begin{equation}\label{NS}
\hat{s}\frac{\partial^2 \bar{u}_i}{\partial x_j\partial x_j}-\frac{\partial \bar{p}}{\partial x_i}=\frac{ \hat{s}^2\bar{u}_i}{\operatorname{Oh}^2},
\end{equation}
where $\hat{p}\equiv \bar{p}/\Sigma=R\tilde{p}/\gamma$ and we have introduced a new rescaled growth rate $\hat{s}\equiv \eta Rs/\gamma=\operatorname{Oh} \bar{s}$. The dispersion relation is obtained by substituting  $\alpha=\sqrt{\hat{s}/\operatorname{Oh}^2+\bar{k}^2}$ and $\beta=1/\hat{s}$ in eq.(\ref{disp}).  The dimensionless control parameters are $\operatorname{Oh}$ and $\bar{r}_0$.
\subsubsection{Stokes flow case}
\begin{figure}[!thbp]
\begin{center}
\includegraphics[height=2.05 in]{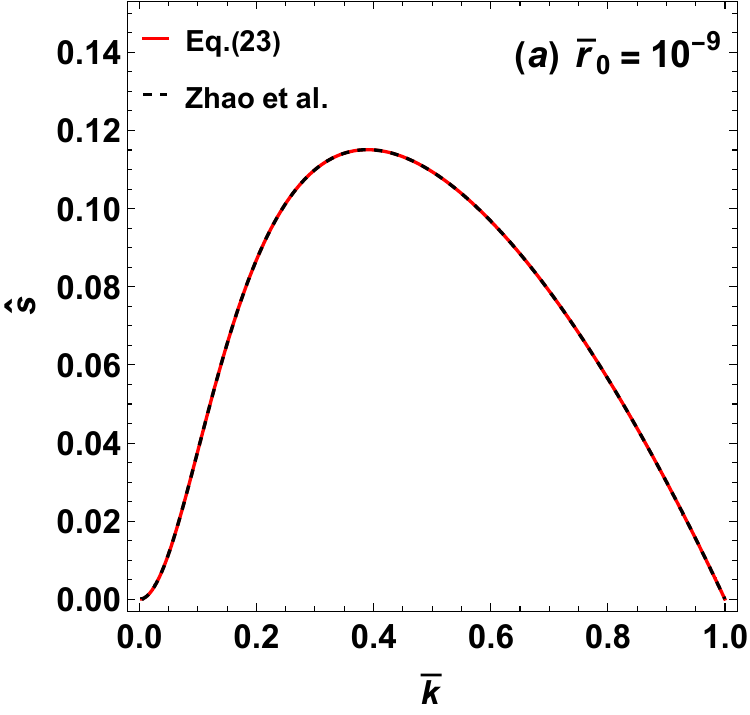}
\includegraphics[height=2.05 in]{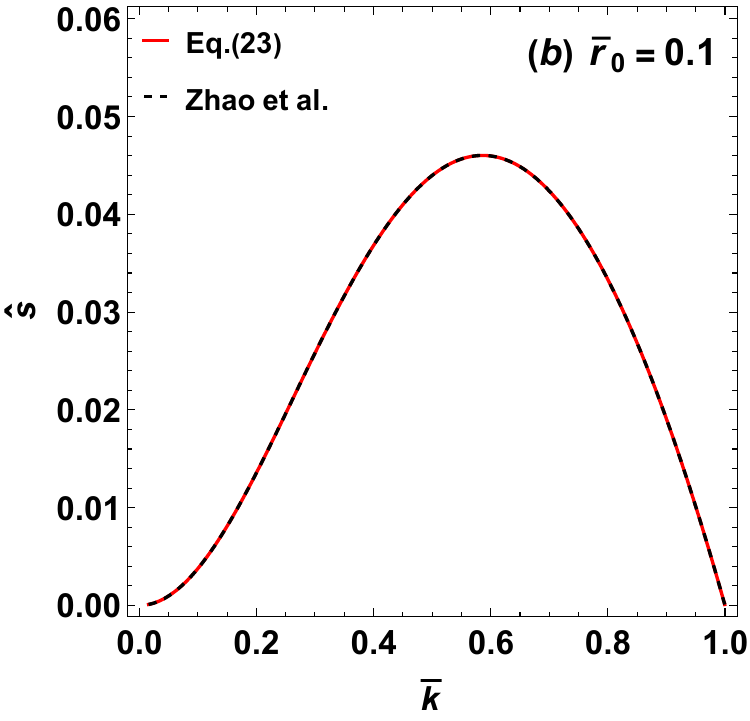}
\includegraphics[height=2.05 in]{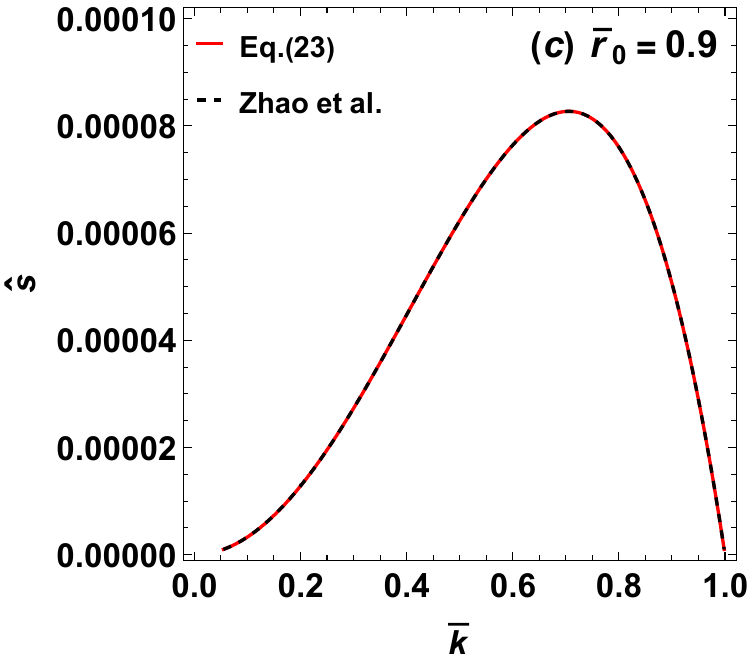}
\caption{The dispersion relation between the growth rate $\hat{s}$ and the wavenumber $\bar{k}$ in the Stokes flow limit for  $\bar{r}_{0}=10^{-9}$ in (a), $\bar{r}_{0}=0.1$ in (b) and $\bar{r}_{0}=0.9$ in (c). Solid lines: our results when substituting $\alpha=\bar{k}$ and $\beta=1/\hat{s}$ for eq. (\ref{disp}). Dashed lines : results from Zhao et al.  \cite{Zhao2023}.} 
\label{fig:6}
\end{center}
\end{figure}

Taking further the limit that $\operatorname{Oh} \rightarrow \infty$, eq. (\ref{NS}) reduces to the Stokes equation in  Laplace space. The remaining dimensionless control parameters is only $\bar{r}_0$. We validate the expression of our dispersion relation (eq. \ref{disp}) in this limiting case by comparing our results  with that from Zhao et al. \cite{Zhao2023} in which the dispersion relation is obtained using the normal mode method to solve the Stokes equations. Fig. \ref{fig:6} shows the comparison for the cases of very small fibre radius ($\bar{r}_0=10^{-9}$), thick coated liquid film ($\bar{r}_0=0.1$) and thin coated liquid film ($\bar{r}_0=0.9$). 
\section*{Conflicts of interest}
``There are no conflicts to declare''.

\section*{Acknowledgements}
The authors gratefully acknowledges financial support from the Research Council of Norway (Project No. 315110). 
\clearpage
\balance
\bibliographystyle{rsc} 

\bibliography{Ridge_form}

\end{document}